\documentclass[aps,
prd,
preprint,
12pt
]{revtex4}

\usepackage{graphicx}
\usepackage{dcolumn}
\usepackage{bm}
\usepackage{color}
\usepackage[normalem]{ulem}

\begin{document}

\title{Sterile neutrinos, dark matter, and resonant effects 
in ultra high energy regimes}

\author{O. G. Miranda$^1$}
\email{omr@fis.cinvestav.mx}

\author{C. A. Moura$^2$}
\email{celio.moura@ufabc.edu.br}

\author{A. Parada$^{1,3}$}
\email{alexander.parada00@usc.edu.co}

\affiliation{$^1$Departamento de F\'{\i}sica, Centro de Investigaci{\'o}n y de
  Estudios Avanzados del IPN, Apdo. Postal 14-740 07000
  M\'exico, D.F., Mexico}

\affiliation{$^2$Universidade Federal do ABC (UFABC), \\ 
Centro de Ci\^encias Naturais e Humanas, \\ 
Rua Santa Ad\'elia, 166, 09210-170 Santo Andr\'e, SP, Brazil}

\affiliation{$^3$Universidad Santiago de Cali (present address), Campus
  Pampalinda, Calle 5  No. 6200, 760001, Santiago de Cali, Colombia}

\date{\today}
\begin{abstract}
Interest in light dark matter candidates has recently increased in the
literature; some of these works consider the role of additional
neutrinos, either active or sterile.
Furthermore, extragalactic neutrinos have been detected with energies
higher than have ever been reported before. This opens a new window of
opportunities to the study of neutrino properties that were
unreachable up to now.
We investigate how an interaction potential between neutrinos and dark
matter might induce a resonant enhancement in the oscillation
probability, an effect that may be tested with future neutrino data.

\end{abstract}

\maketitle

\section{Introduction}
It is well known that neutrinos propagating through a material medium
experience an enhancement effect in the oscillation probability, the
so-called MSW effect~\cite{Wolfenstein:1977ue}. The standard MSW
effect takes into account the interaction of active neutrinos with
electrons and quarks.

However, the oscillation of standard flavor neutrinos into non
standard sterile neutrinos has been considered in previous
works~\cite{Lunardini:2000swa} as, e.g., an explanation for the
reactor neutrino anomaly~\cite{Mention:2011rk}, for the LSND and
MiniBooNE experimental results~\cite{Conrad:2013mka}, in the context
of primordial
nucleosynthesis~\cite{Kainulainen:1990ds,Enqvist:1990ad,Barbieri:1990vx,Enqvist:1991qj,Foot:1995bm,Dolgov:2000jw},
and in supernovae~\cite{Wu:2013gxa}.  In addition, sterile neutrinos
appear in models attempting to explain the dark matter
problem~\cite{Merle:2013gea}, either as the main component for the
dark matter content or as an additional subleading component of a
multiparticle dark matter model.  Couplings between neutrino, either
active or sterile, and dark matter have also been studied in many
different
contexts~\cite{Huang:2013zga,Aarssen:2012fx,Mangano:2006mp,Fayet:2007ua,Boehm:2003hm,Fox:2008kb,Berezhiani:1995yi,Berezhiani:1995am,Laha:2013xua,Shoemaker:2013tda,Dasgupta:2013zpn,Hannestad:2013ana}.

We propose that, if there is a mixing between active and sterile
neutrinos, high-energy neutrinos interacting with dark matter may
suffer a kind of MSW effect when they propagate in a dark matter
medium.  We show that if there is an interaction of neutrinos with
dark matter, their corresponding potential might induce a resonant
effect, in just the same way as active neutrinos are affected by the
interaction with the electrons of a medium.

\section{Dark matter and resonant effects}

We begin our analysis by showing the neutrino evolution equation, which 
includes both ordinary and dark matter potentials. We study a
simplified picture with one sterile neutrino, $\nu_s$, and an active
one, $\nu_{\alpha}\,$. For a neutrino energy, $E$, the
evolution equation can be written as

\begin{equation}
i \frac{\rm d}{{\rm d}t}
\left(\begin{array}{c}
\nu_\alpha \\
\nu_s \\
\end{array} \right)
=
{\rm M}_\alpha
\left( \begin{array}{c}
\nu_\alpha \\
\nu_{s} \\ \end{array} \right) \,,
\label{eq:1}
\end{equation}
with
\begin{equation}
{\rm M}_\alpha=\left(\begin{array}{cc} 
-\frac{\Delta m^{2}_{i4}}{4E}\cos2\theta_{0} + V_{\nu_\alpha f} + V_{\nu_\alpha \chi} & 
\frac{\Delta m^{2}_{i4}}{4E}\sin2\theta_{0} \\ 
& \\ 
\frac{\Delta m^{2}_{i4}}{4E} \sin2\theta_{0} & 
\frac{\Delta m^{2}_{i4}}{4E}\cos2\theta_{0} +  V_{\nu_s \chi} \\
 \end{array} \right) \, ,
\label{eq:msw-matrix}
\end{equation}
where $\Delta m^{2}_{i4} = m^{2}_{4}-m^{2}_{i}$, and
the angle $\theta_{0}$ is the vacuum mixing angle between the
sterile  and  the active neutrino; $V_{\nu_\alpha f}=
V_{\nu_\alpha f}^{CC} + V_{\nu_\alpha f}^{NC} $ accounts for the
well-known interaction potential of the active neutrino with ordinary
fermions; $V_{\nu_\alpha \chi}$ takes into account the potential due
to a possible interaction between active neutrinos and dark matter.
In this work, we also investigate the effect of the potential $V_{\nu_s \chi}$,
coming from the interaction of sterile neutrinos with dark
matter. This interaction naturally appears in different extensions of
the Standard Model, where many dark particles, including sterile
neutrinos, could populate the dark sector and interact among
themselves~\cite{Berezhiani:1995yi,Berezhiani:1995am,Zhang:2013ama}.
The interaction potential  $V_{\nu_s f}$ has already been studied~\cite{Bramante:2011uu} and is negligible compared to the other potentials in Eq.~(\ref{eq:msw-matrix}). Therefore, we do not include it in our calculations. 

The resonance condition derived from Eq. (\ref{eq:1}) is then given by 
\begin{equation}
\Delta m^{2}_{i4}\cos2\theta_{0} = 2E (V_{\nu_\alpha f} + V_{\nu_\alpha \chi} - V_{\nu_s \chi}) \,.
\end{equation}

We can write these potentials as follows:
\begin{eqnarray}
\centering
\label{eq:4}
V_{\nu_\alpha f}  &=& \frac{1}{4}\frac{g^2}{m_W^2}(N_\alpha - N_n/2) = \sqrt2 G_F (N_\alpha - N_n/2) \,; \\
V_{\nu_\alpha \chi} & \sim & \frac{g_{\nu_\alpha} g_\chi}{m_I^2} N_\chi = G_{\nu_\alpha}'N_\chi=\varepsilon_{{\nu_\alpha}\chi}G_FN_\chi \,; \\
V_{\nu_s \chi} & \sim & \frac{g_{\nu_s} g_\chi}{m_I^2} N_\chi = G_{\nu_s}'N_\chi=\varepsilon_{{\nu_s}\chi}G_FN_\chi\,,
\end{eqnarray}
where $N_\alpha$, $N_n$, and $N_\chi$ are, respectively, the number
density of leptons, neutrons, and dark matter particles interacting
with neutrinos.  In Eq.~(\ref{eq:4}), $g$ is the Standard Model
coupling constant and $m_W$ is the $W$~boson mass; while,
$g_{\nu_\alpha}$, $g_{\nu_s}$, and $g_\chi$ represent the coupling
constants of the corresponding particle (active neutrino, sterile
neutrino, and dark matter) with an intermediate gauge boson with mass
$m_I$. The parameters $\varepsilon_{\nu_{\alpha,s}\chi}$ account for
the coupling strength in terms of the Fermi constant $G_F$.

Using the above expressions for the potentials, the resonance condition
is written as
\begin{equation}
\Delta m^{2}_{i4}\cos2\theta_{0} = 2EG_F[\sqrt2(N_\alpha - N_n/2) + (\varepsilon_{{\nu_\alpha}\chi}-\varepsilon_{{\nu_s}\chi})N_\chi] \,.
\label{resonance}
\end{equation}

The standard contribution to this equation, $V_{\nu_\alpha f} = \sqrt2
G_F (N_\alpha - N_n/2) $, is zero for the case of electron neutrinos,
considering an astrophysical environment with $N_e \approx N_n/2$, and
$V_{\nu_{\mu,\tau} f} =-\sqrt2 G_F N_n/2$ for muon and tau neutrinos
($N_\mu\approx N_\tau\approx0$). In practice, $V_{\nu_{\mu,
      \tau}f}$ will be negligible in comparison with the new
  contributions from $V_{{\nu_\alpha}\chi}$ and $V_{{\nu_s}\chi}$ and,
  therefore, our results will apply to any of the three active
  neutrino species.  For the estimate of the dark matter number
density, $N_\chi$, we consider that the main contribution arise from a
single heavy dark matter particle with mass $m_\chi$ and, therefore,
the relevant density in our case will take the value $N_\chi =
\rho_{\chi} / m_\chi$~\footnote{It is possible that in some models the
  heaviest dark particle is different from the particles interacting
  with the sterile neutrinos, however, we would expect the number
  density to be approximately equal; an example in this direction
  could be a mirror model where the interacting particle is a mirror
  electron and the heaviest dark matter particle is a mirror proton.}.
To estimate $\varepsilon_{\nu_{\alpha,s}\chi}$ we need to study in detail 
the coupling constants 
$g_{\nu_\alpha,\nu_s,\chi}$.  Recently, the interest in models with an
intermediate
boson with a relatively light mass $m_I$ has grown,
especially in the context of the dark matter
problem~\cite{Aarssen:2012fx,Mangano:2006mp,Fayet:2007ua,Boehm:2003hm,Fox:2008kb,Laha:2013xua}.
We show in Table~\ref{table:models} an incomplete list of 
values for $\varepsilon_{\nu_{\alpha,s}\chi}$ in these types of models.
Notice that the coupling of active
neutrinos with dark matter can be strongly constrained ($g_\chi g_\nu\sim
10^{-6}$~\cite{Fayet:2007ua,Boehm:2003hm}) while for the sterile case the 
constraints are weaker, as should be expected.

\begin{table}
 \begin{tabular}{lcccc}
\hline \hline 
Ref. & $\frac{(g_\chi)(g_\nu)}{(m_I/[{\rm MeV}])^2}$ & $\varepsilon_{\nu_{e}\chi}$ & $\varepsilon_{\nu_{s}\chi}$ & 
$\frac{m_\chi}{[{\rm MeV}]}$ \\ 
\hline\hline 
Aarssen {\it et al.}~\cite{Aarssen:2012fx}   & $\frac{(0.7)(10^{-6}-10^{-1})}{10^{-2}-1}$ & 0  & $10^{5}-10^{15}$ & $10^{6}$ \\ 
Mirror~\cite{Berezhiani:1995yi,Berezhiani:1995am} &
$\frac{(1)(1)}{(30 m_W)^2}$ & 0 & $10^{-3}$ & $10^{3}$ \\ 
Fayet~\cite{Fayet:2007ua,Boehm:2003hm} & $<\frac{10^{-6}}{1}$ & $<10^5$ & 0 & $10$ \\ 
Mangano {\it et al.}~\cite{Mangano:2006mp} & $<\frac{10^{-3}}{1}$ & $<10^8$& 0 & $10$ \\
\hline\hline
\end{tabular}
\caption{Coupling constants and mass estimates from different models. }
\label{table:models}
\end{table}

\section{An application}

We now turn our attention to the search for physical processes that
could be sensitive to the effects of the $\nu-\chi$ interaction
potential, trying to shed some light in the study of two hidden
sectors: sterile neutrino and dark matter sectors.

\begin{figure}
\includegraphics[width=0.6\columnwidth]{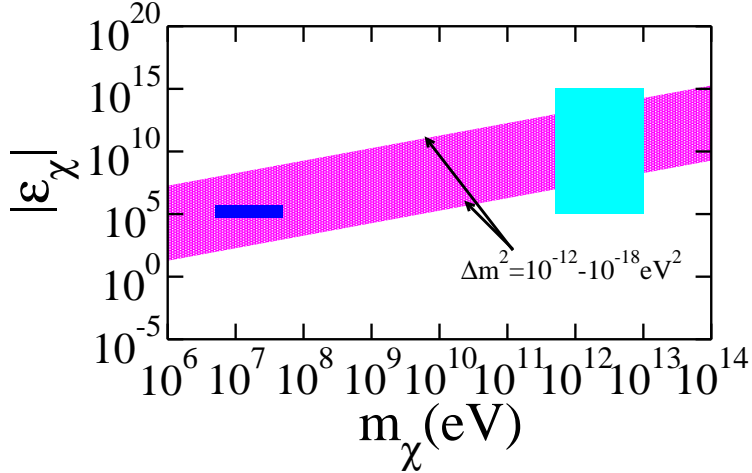}
\caption{Coupling strength $|\varepsilon_\chi|$ and dark matter mass
  $m_\chi$ corresponding to oscillation resonance for active neutrinos
  with energy $E=10^{15}$~eV and $\Delta m^2$ in the range from
  $10^{-18}$~eV$^2$ up to $10^{-12}$~eV$^2$, propagating in the
  vicinity of our galaxy.  In this plot we have considered the limit
  case of a vanishing mixing angle $\theta_0$. As a matter of
  comparison, we plot different models considered in the
  literature. In the left dark blue box are those of
  Fayet~\cite{Fayet:2007ua,Boehm:2003hm} and in the right light blue
  box is Aarssen {\it et al.}~\cite{Aarssen:2012fx}.}\label{fig:name}
\end{figure}

In order to find the conditions in which an oscillation resonance can
take place according to Eq.~(\ref{resonance}), we compute the values
of $\Delta m^2$ (from now on we omit the subscript $i4$ from $\Delta
m^2_{i4}$),
$\varepsilon_\chi=(\varepsilon_{{\nu_\alpha}\chi}-\varepsilon_{{\nu_s}\chi})$,
and $m_\chi$ that induce such an effect.  We present our results in
Fig.~(\ref{fig:name}) for the parameter space $|\varepsilon_\chi|$ vs
$m_\chi$ in a range of $\Delta m^2$ values, taking as a first
approximation $\theta_{0}\approx0$. Notice that $\varepsilon_\chi$
changes sign depending on which coupling is stronger: whether it is
$\varepsilon_{{\nu_\alpha}\chi}$ or $\varepsilon_{{\nu_s}\chi}$.
Therefore, the resonance condition is valid only for neutrinos (if
$\varepsilon_{{\nu_\alpha}\chi} > \varepsilon_{{\nu_s}\chi}$) or for
antineutrinos (if $\varepsilon_{{\nu_s}\chi} >
\varepsilon_{{\nu_\alpha}\chi}$).

We have conducted an analysis considering the dark matter around our
galactic halo,
where, on average, the electron density can be approximately 
$N_{e}=3\times 10^{-16}$~eV$^{3}$~\cite{deAvillez:2012ue} and 
it
is expected that $\rho_{\chi} = 0.3\, {\rm GeV} \cdot {\rm cm}^{-3}$~\cite{Foot:2012rk}.  
We compute our result for a fixed neutrino energy of $E=10^{15}$~eV.
In particular, we show a tilted band that corresponds to the range
$10^{-18}$~eV$^2 < \Delta m^2 < 10^{-12}$~eV$^2$, that was previously
studied in a similar context, although for pseudo-Dirac
oscillations~\cite{Esmaili:2012ac,Joshipura:2013yba,Esmaili:2009fk,Beacom:2003eu,Keranen:2003xd,Crocker:2001zs}.
In the same Fig.~(\ref{fig:name}), we also plot the space of
parameters $|\varepsilon_\chi|$-$m_\chi$ obtained from the work of
Aarssen {\it et al.}~\cite{Aarssen:2012fx} (light blue box on the
right). These authors discussed the possibility of an interaction
between dark matter and neutrinos in order to address $\Lambda$CDM
small-scale problems.  We consider the couplings discussed in this
article as a guidance for a sterile neutrino coupling with dark
matter. Finally, for the interaction between active neutrinos and dark
matter, we plot the space of parameters (dark blue box on the left)
constrained in Ref.~\cite{Fayet:2007ua,Boehm:2003hm}.

It is quite interesting that this $\Delta m^2$ range is consistent
with an oscillation resonance for coupling constants and masses for
dark matter candidates proposed in other articles, especially because
it has already been noticed that a pseudo-Dirac oscillation could lead to
an UHE neutrino flux deficit~\cite{Esmaili:2012ac}.  If the dark
matter surrounding our galaxy induces such a resonance effect there
could be an energy range where active neutrinos convert to sterile
neutrinos. This would cause an important change in the neutrino flux
spectrum, as long as the source is extragalactic.

Several constraints on the UHE neutrino flux, coming from
Auger~\cite{Abreu:2011pf} and ANTARES~\cite{Biagi:2011kg} have been
reported, and a bound on neutrinos from gamma ray bursts has also been
presented by Icecube~\cite{Abbasi:2012zw}.
Recently, Icecube also reported the detection of neutrinos coming from
extraterrestrial sources: The first report presented the detection of
two electron neutrino events with energies around
PeV~\cite{Aartsen:2013bka}, while a later report presented data on the
detection of 26 neutrino events in the range of
$30-300$~TeV~\cite{Aartsen:2013jdh}. The detection of 37 events in
three years of data collection was presented
in~\cite{Aartsen:2014gkd}.  Additional research is needed in order to
develop a more complete understanding of this
data~\cite{Halzen:2013dva}.

With the accumulation of data from IceCube,
Auger, and future telescopes as the KM3Net, we would have a better 
understanding of the galactic and extragalactic neutrino spectrum.
In this context, we would like to study if the interaction potential
between neutrino and dark matter that we propose might induce an
oscillation resonance in the UHE regime. If the experiments collect
sufficient data, it might be possible to observe the MSW mechanism for
dark matter as a distortion in the UHE neutrino spectrum. 

Additionally, instead of considering the limit of a vanishing
mixing angle, we compute the survival probability for active
neutrinos as:

\begin{equation}
P(\nu_{\alpha}\to\nu_{\alpha}) = 1 - \sin^{2}(2\theta_{m})\sin^{2}
\left(\pi\frac{L^{osc}_{0}}{L^{osc}_{m}}\right)\,.
\end{equation}
In this expression
\begin{equation}
\sin^{2}(2\theta_{m}) = \frac{\sin^{2}(2\theta_{0})}
{\cos^{2}(2\theta_{0})\left(1 - \frac{( V_{\nu_{\alpha}\chi} - V_{\nu_{s}\chi})}{V_{R}}\right)^{2} + \sin^{2}(2\theta_{0})}\,,
\end{equation}
where $V_{R} = \frac{\Delta m^{2}}{2E}\cos(2\theta_{0})$. And the  oscillation length in 
matter is given by 
\begin{equation}
L^{osc}_{m} = \frac{L^{osc}_{0}}{\sqrt{\cos^{2}(2\theta_{0})\left(1 - 
\frac{ V_{\nu_{\alpha}\chi} - V_{\nu_{s}\chi})}{V_{R}}\right)^{2} + \sin^{2}(2\theta_{0})}}\,.
\end{equation}

\begin{figure}[h!]
\includegraphics[width=0.6\textwidth]{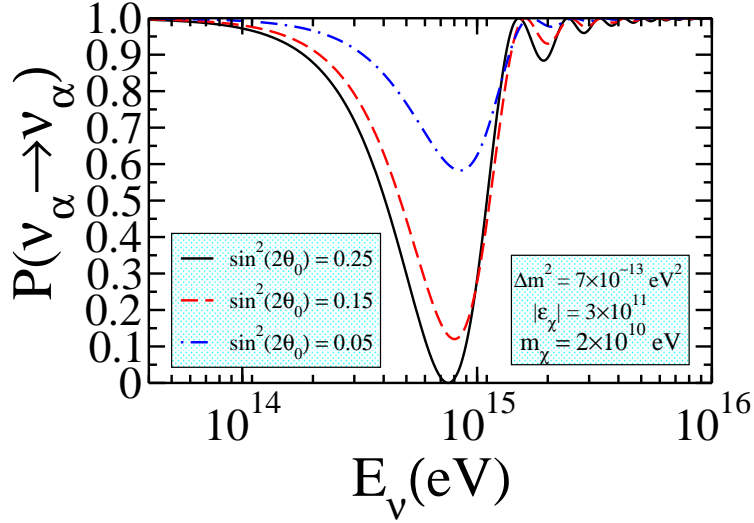}
\caption{Survival probability $P(\nu_{\alpha}\to\nu_{\alpha})$
 as a function of the neutrino energy $E_\nu$, considering the galactic halo
 average dark matter density. }
\label{surv_prob}
\end{figure}

We have computed the survival probability, for different values of
$\sin^2(2\theta_0)$, for the case in which the neutrino squared mass
difference is given by $\Delta m^2 = 7\times 10^{-13}$~eV$^2$, with a
coupling $|\varepsilon_\chi|=3\times 10^{11}$, and a dark matter
mass $m_\chi=2\times 10^{10}$~eV. We found the
resonant energy around $E=8\times 10^{14}$~eV, as shown
in Fig.(\ref{surv_prob}). From this figure we
see that a mixing of the order $\sin^2(2\theta_{0})=0.25$
could give a maximal conversion with a wide energy window.
These values also make an effective oscillation length possible in
conformity with the expected dark matter halo
dimension~\cite{Diemand:2009bm}, as the oscillation length in dark
matter may be given by $L^{osc}_{m} = \frac{4\pi
  E}{\sin(2\theta_{0})\Delta m^{2}} \sim 10^{18}$~km.

Although this suggests that the high energy spectrum of extragalactic
neutrinos could be affected by the existence of sterile neutrino and
its interaction with dark matter, a more detailed study must be
conducted. For instance, in order to have an MSW resonance, other
conditions must be fulfilled~\cite{Lunardini:2000swa}, like the
adiabaticity condition. We start from the definition of the
adiabaticity parameter~\cite{Giunti:2007ry}:

\begin{equation}
\gamma=\frac{(\Delta m^2_{\rm{m}})^2}{2E\sin(2\theta_m)|dA_{cc}/dr|}\,,
\end{equation}
where
\begin{equation}
\sin(2\theta_m)=\Delta m^2\sin2\theta_0/\Delta m_{\rm m}^2\,,
\end{equation}
\begin{equation}
\Delta m_{\rm m}^2=\sqrt{(\Delta m^2\cos2\theta_0-A_{cc})^2+(\Delta m^2\sin2\theta_0)^2}\,,
\end{equation}
and, in our case,
\begin{equation}
A_{cc}=2E(V_{\nu_\alpha f} + V_{\nu_\alpha \chi} - V_{\nu_s \chi}) 
 \simeq 2E\varepsilon_\chi G_F N_\chi\,.
\end{equation}
It is possible to note that the adiabaticity condition can be expressed as
\begin{equation}
\gamma=\frac{((\Delta m^2\cos2\theta_0-2E\varepsilon_\chi G_F N_\chi)^2+(\Delta m^2\sin2\theta_0)^2)^{3/2}}{4E^2\Delta m^2 \sin2\theta_0 \varepsilon_\chi G_F |dN_\chi/dr|} >>1\,.
\end{equation}
This condition is satisfied for the case of a constant density dark
matter distribution.  Another important condition to be fulfilled, in
order to have significant conversion probability, is that the width
$d$ of dark matter~\cite{Lunardini:2000swa}, should be larger than a
minimum width $d_{min}$.  Following closely
reference~\cite{Lunardini:2000swa}, in our analysis, this condition is
given by
\begin{equation}
d = \int N_{\chi}(L)dL \geq d_{min} = \frac{1}{|\varepsilon_{\chi}|G_{F}\tan 2\theta_{0}}\,,
\end{equation}
where $L$ denotes the distance travelled by the neutrino in the dark
matter medium. Taking into account the parameters considered for
Fig.~(\ref{surv_prob}), that is $|\varepsilon_\chi| = 3\times10^{11}$ and
$\sin^2(2\theta_0)=0.25$, we obtain $d_{min} = 1.3\times 10^{21}~$cm$^{-2}$, 
while, for a dark matter halo of $6\times10^{18}$~km, 
$d = 9\times 10^{21}~$cm$^{-2}$. This shows that the
width of dark matter is approximately one order of magnitude bigger
than the minimum width, making the conversion from active to sterile
neutrinos possible.  

Although it is promising that these resonance
  conditions~\cite{Lunardini:2000swa} are satisfied for a constant
  distribution, it would be necessary to study the case of a more
  realistic dark matter profile. 
 In order to obtain a first
  estimate, we consider a halo density of the form
\begin{equation}
\rho(r) = \frac{\rho_0}{(r/R)^\delta[1 + (r/R)^\alpha]^{(\beta-\delta)/\alpha}}\,,
\end{equation}
where $\alpha$, $\beta$, $\delta$, and $R$ (in kpc) depend on the
specific model to be considered~\cite{Bertone:2004pz}. We have
  computed the adiabaticity parameter for the widely known profiles of
  Navarro, Frenk, White~\cite{Navarro:1995iw} ($\alpha = 1$, $\beta =
  3$, $\delta = 1$ and $R=20$~kpc), Kravtsov
  et. al.~\cite{Kravtsov:1997dp} ($\alpha = 2$, $\beta = 3$, $\delta =
  0.4$ and $R=10$~kpc), Moore et. al.~\cite{Moore:1999gc} ($\alpha =
  1.5$, $\beta = 3$, $\delta = 1.5$ and $R=28$~kpc), and for the
  modified isothermal profile~\cite{Bergstrom:1997fj} ($\alpha = 2$,
  $\beta = 2$, $\delta = 0$ and $R=3.5$~kpc).  We found that the
  three resonance conditions are satisfied for all the profiles
 if we consider a dark
  matter mass of the order of $100$~MeV, the same parameter for
  neutrino mass difference, $\Delta m^2 = 7\times 10^{-13}$~eV~$^2$,
  the coupling $|\varepsilon_\chi|=3\times 10^{11}$, and a neutrino energy
  $E=10$~TeV. For these parameters, in the case of the Navarro, Frenk
  and White profile, the minimum value of the adiabaticity parameter
  is $\gamma \approx 13$ and the resonance is located
    around $18$ kpc from the galactic center. The dark matter width 
    is $d=1.3\times10^{23}$~cm$^{-2}>d_{min}$. These results are
  encouraging and suggest that a wider region of parameters,
  satisfying the resonance conditions, could be found by conducting a
  detailed study for these profiles.

\section{Conclusions}

In summary, in this work we have studied the possibility that
neutrinos might have a resonant effect in the presence of additional 
sterile neutrino states and dark matter. We have conducted an analysis
of the necessary couplings of dark matter with either active or
sterile neutrinos in order to have such an effect.
Our results show that, if the phenomenological models discussed here
happen in nature, they may induce a resonant oscillation of high
energy active to sterile neutrinos.  We have shown values of $\Delta
m^2_{i4}$ where there could be a resonant effect for an adequate range of
neutrino couplings and dark matter mass.
The mechanism discussed here could be tested with future ultra high
energy neutrino data, for instance, from the IceCube experiment.

\begin{acknowledgments}
  This work has been supported by EPLANET, CONACyT grant 166639, and
  SNI-Mexico.  C.A.M. thanks the CINVESTAV Physics Department for
  their hospitality while conducting a portion of this study during
  his invited visit. 
The work of C.A.M. was partially supported by Fundacao de Amparo a
Pesquisa do Estado de Sao Paulo (FAPESP), under the grant 2013/22079-8
The work of A.P. was partially supported by Universidad Santiago de Cali
under grant 935-621114-031.
We would like to thank the referee for very
  useful comments on the manuscript.
\end{acknowledgments}

\end{document}